# Virtual impactor-based label-free bio-aerosol detection using holography and deep learning

*Yi Luo*[1,2,3,+]  e-mail: yluo2016@ucla.edu
*Yijie Zhang*[1,2,3,+]  e-mail: yijiezhang@g.ucla.edu
*Tairan Liu*[1,2,3]  e-mail: liutr@ucla.edu
*Alan Yu*[1,4]  e-mail: alanyu111@g.ucla.edu
*Yichen Wu*[1,2,3]  e-mail: wuyichen@g.ucla.edu
*Aydogan Ozcan*[1,2,3*]  e-mail: ozcan@ucla.edu

[1]Electrical and Computer Engineering Department, University of California, Los Angeles, California 90095, USA

[2]Bioengineering Department, University of California, Los Angeles, California 90095, USA

[3]California Nano Systems Institute (CNSI), University of California, Los Angeles, California 90095, USA

[4]Computer Science Department, University of California, Los Angeles, California 90095, USA

[+]Equal contributing authors

[*]Correspondence: Prof. Aydogan Ozcan

E-mail: ozcan@ucla.edu

# Abstract

Exposure to bio-aerosols such as mold spores and pollen can lead to adverse health effects. There is a need for a portable and cost-effective device for long-term monitoring and quantification of various bio-aerosols. To address this need, we present a mobile and cost-effective label-free bio-aerosol sensor that takes holographic images of flowing particulate matter concentrated by a virtual impactor, which selectively slows down and guides particles larger than 6μm to fly through an imaging window. The flowing particles are illuminated by a pulsed laser diode, casting their inline holograms on a CMOS image sensor in a lens-free mobile imaging device. The illumination contains three short pulses with a negligible shift of the flowing particle within one pulse, and triplicate holograms of the same particle are recorded at a single frame before it exits the imaging field-of-view, revealing different perspectives of each particle. The particles within the virtual impactor are localized through a differential detection scheme, and a deep neural network classifies the aerosol type in a label-free manner, based on the acquired holographic images. We demonstrated the success of this mobile bio-aerosol detector with a virtual impactor using different types of pollen (i.e., bermuda, elm, oak, pine, sycamore, and wheat) and achieved a blind classification accuracy of 92.91%. This mobile and cost-effective device weighs ~700 g and can be used for label-free sensing and quantification of various bio-aerosols over extended periods since it is based on a cartridge-free virtual impactor that does not capture or immobilize particulate matter.

# Introduction

Bio-aerosols account for 5-34% of indoor particulate matter (PM)[1]. They are airborne microparticles originating from plants, animals, and living or dead microorganisms[2]. Bio-aerosols can easily enter the respiratory tract during inhalation due to their microscopic size. Exposure to bio-aerosols has been related to a wide range of health issues[3,4]. Some bio-aerosols cause irritation and allergic reactions, such as pollen[5,6]; some others, like fungal and bacterial[7,8] PM, can spread infectious and respiratory diseases. For example, bio-aerosols served as an important transmission route during the COVID-19 pandemic[9,10]. They may also lead to an increased risk of cancer[4,11,12]. Conventional sensing of bio-aerosols includes two steps: the aerosols are first sampled using, e.g., an impinger, a cyclone, an impactor, or a filter[13,14], and then analyzed in a central lab under a microscope with fluorescence labeling or through a culture-based procedure by a microbiology expert[15]. Other technologies, such as polymerase chain reaction (PCR)[16] and enzyme-linked immunosorbent assays (ELISA)[17], are also applied to better identify the captured bio-aerosols with high sensitivity and specificity. However, the complicated procedures and the need for well-trained experts hinder their widespread use for continuous monitoring of human exposure to bio-aerosols.

For field-portable bio-aerosol monitoring devices that integrate aerosol sampling and inspection, a major challenge is identifying the collected particles. Most devices avoid this challenge by selectively sampling a few types of bio-aerosols using specific antibodies. This antibody-antigen specific reaction can be sensed using different mechanisms, such as lateral flow-based immunoassays[18], vibrational cantilevers[19], surface plasmon resonance-based sensors[20], or Raman spectroscopy[21]. However, the immunoreaction limits the collection efficiency and throughput of the device, and non-specific binding events can cause false-positive detections. Also, the utilization of antibodies creates storage and shelf-life-time issues for these sensors, and it lacks scalability to cover a larger variety of bio-aerosols that might be present in different parts of the world during different seasons. In another embodiment, analyzing the autofluorescence signals of individual bio-aerosols excited by ultraviolet (UV) light was utilized as a label-free method for bio-aerosol detection[22,23], but this approach suffered from low specificity due to the insufficient information provided by the weak autofluorescence signals. As an alternative approach, a field-portable, cost-effective platform for high-throughput quantification of aerosols using mobile microscopy was also reported[24,25]. This device incorporates an impaction-based aerosol sampling method: a high-speed airstream carries particles moving from the impactor nozzle to a transparent polymer substrate that faces the inlet flow direction. Large aerosols detach from the streamline due to their inertia and are physically collected/captured by the substrate. However, the transparent impactor used for particle collection in this platform suffers from an overfill of the sampling substrate: excessive particles captured on the polymer substrate occlude the imaging field-of-view, preventing new particles from being detected. Therefore, the impactor cartridge must be frequently replaced, which makes the platform inadequate for long-term unsupervised operation.

A good substitution for a physical impactor can be a virtual impactor, which replaces the collection substrate of an impactor with a middle channel (a collection probe)[26], where only a minor portion of the input flow will go through. Most of the input air leaves the device via by-pass channels, leading to a sharp flow direction change, where large particles detach from the major streamline and enter the middle channel, as their greater inertia prevents them from following the drastic flow direction change. Consequently, particles are separated based on their inertia, and large ones are concentrated inside the middle channel. Virtual impactors have been widely implemented for ambient fine particle sensing[27–29], especially monitoring PM in air[30–35]. Recent efforts have also utilized this platform for bio-aerosol detection in indoor environments[36,37]. However, to classify the type of the collected particles for bio-aerosol sensing, the airflow from the middle channel needs to go through a filtration step, where the flowing bio-aerosols are transferred to a physical filter[37] or a bio-aerosol collector (an impinger)[38]. Further analytical examination steps, such as culturing and PCR analysis[39], are applied to reveal the species of the collected bio-aerosols.

In this work, we present a virtual impactor-based, cartridge- or filter-free bio-aerosol detection method that combines computational imaging and deep learning to sense and classify bio-aerosols without any external labels or chemical sample processing steps (Fig. 1(a)). In this mobile and cost-effective device, a virtual impactor was designed and 3D printed to concentrate the flowing particles larger than ~6 μm, which covers the size range of most pollen species. An imaging window with a sensing volume of 25 mm$^3$ was placed on the middle channel, and inline lensfree holography was used to image the passing-by particles, owing to its capability of recording volumetric information with a large field-of-view[40–46]. Coherent illumination from a laser diode shined through the imaging window, forming inline holograms on a complementary metal-oxide semiconductor (CMOS) image sensor placed on the other side, right next to the imaging window (Fig. 1(b) and (c)). Three illumination pulses were fired during each frame, casting three different holographic images of each particle – all captured at the same frame. With each illumination pulse, a clear lensfree hologram of the particle that is free from motion blur and rolling shutter artifacts was captured by the CMOS image sensor. After three pulses, the CMOS image sensor integrated all three holograms of the same particle at three different lateral locations, forming a unique triplicate holographic pattern per flowing particle (Fig. 1(d)). The particles are localized by a differential detection algorithm, and a trained deep neural network was used to classify the pollen type of each particle from its auto-focused holograms. As a demonstration of the proof of concept of this mobile and cost-effective system, we imaged aerosols of six different types of pollen: bermuda, elm, oak, pine, sycamore and wheat, which are widespread in North America and Europe[47,48]. With the triplicate holographic patterns per particle, a majority voting was applied to the classification decision of each particle, achieving a pollen classification accuracy of 92.91% (Fig. 1(e)).

To the best of our knowledge, this is the first demonstration of an imaging-based virtual impactor design that enables label-free bio-aerosol detection using neural networks, without the need for any filtration or chemical processing. This device is compact, cost-effective and light-weight (~700 g) and since it does not require a cartridge or filter for bio-aerosol sensing, it enables air quality monitoring over an extended period of time without any supervision. This AI-based bio-aerosol detection and classification device provides a unique solution to indoor air quality monitoring and label-free bio-aerosol sensing.

## Results

### Portable bio-aerosol sampling device using a virtual impactor

Our portable device designed to sample and image bio-aerosols contains three major parts: a virtual impactor that collects and slows down the flowing pollens in its middle channel, a lens-free holographic imaging system capturing microscopic images of pollens, and a controlling circuit automating the entire workflow. The virtual impactor contains one middle channel with a designed flow rate of 10 mL/min, and two symmetrical by-pass channels with a total flow rate of 1L/min. The physical dimensions of the virtual impactor were optimized by finite element method (FEM) simulations using COMSOL (Figs. 2(a) and (b)). Input aerosols larger than 6.3 μm (cut-off diameter) have more than 50% probability of entering the virtual impactor through the middle channel, while other smaller particles leave via by-pass channels (Fig. 2(c)). Based on this, the pollen particles to be detected and classified are concentrated in the middle channel, where the flow rate is ~100 times smaller than the input flow rate. Consequently, each pollen type of interest has a significantly higher concentration inside the middle channel than in ambient air. Two fans are used to power the flow inside the middle and by-pass channels independently, whose flow rates are monitored by two separate flow meters. The real-time signals are sent to an Arduino microcontroller, where a simple PID feedback loop was implemented to adjust the fans' speed (see Fig. 2(d) and Supplementary Fig. S1).

An imaging window with a volume of 5×5×1 mm$^3$ is opened at the center of the middle channel (Fig. 2(b)), sealed by coverslip glasses. A CMOS image sensor to capture the lensfree holograms of the particles flowing over the imaging window was placed next to the channel, touching the coverslip glass. The axial distance from the bottom of the channel to the CMOS sensor is ~1.5 mm. A customized Graphical User Interface (GUI) was designed to control the CMOS imager and take holographic videos at a frame rate of 3.5 fps. If CW illumination were to be used to form holograms, the high speed of the concentrated particles (~0.17m/s) would normally introduce a strong motion blur and rolling shutter artifacts. To capture motion blur-free and undistorted holograms, the imaging system was configured to mimic strobe photography (Fig. 1(d)). When all the pixels of the CMOS are turned on to collect photons, a signal was sent to our customized pulse generation circuit, firing a train of three successive pulses using a laser diode ($\lambda$=515nm) that shines through the imaging window. Under each laser pulse, all the flowing particles above the imaging window form their holograms captured by the CMOS image sensor. The pulse duration was short enough (6.9 μs) to avoid motion blur for each holographic pattern. The particles travel along the flow direction during the interval between two pulses ($\delta t = 699 \mu s$). As a result, after the pulse train, three individual holograms were acquired by the CMOS image sensor, through a *single frame* with triplicate holographic patterns for each pollen particle flowing inside the middle channel. Each captured hologram was transferred into the controlling laptop for further processing. The entire mobile device has dimensions of 24.1 cm×10.5 cm×10.2 cm and weighs ~700g. The components of our prototype cost ~$923 to build, and its detailed list can be found in Table 1.

Table 1. Purchase costs of the components used in our mobile bio-aerosol sensor prototype.

| Components | Cost ($) |
|---|---|
| 515 nm laser diode | $15.36 |
| IDS-UI-3592LE-C-VU CMOS sensor | $608.00 |
| AWM3100V flow sensor | $111.00 |
| FS2012 flow meter | $41.25 |
| 9GA0312P3K0011 DC fan | $15.34 |
| 9CRB0412P5S201 DC fan | $30.09 |
| Customized PCB board | $20.00 |
| AC/DC wall mount adaptor | $16.87 |
| Arduino Nano Every | $11.85 |
| TLC5916 LED driver | $1.36 |
| Mechanical supporting parts | $2.00 |
| 3D printing material | $50.00 |
| **Total** | **$923.12** |

## Label-free bio-aerosol imaging and classification

To demonstrate the proof-of-concept of our mobile device for label-free bio-aerosol detection, we targeted aerosols containing six different types of pollens: bermuda, elm, oak, pine, sycamore, and wheat. Purified aerosols containing only a single pollen type were generated using a customized particle generator that directly connects to our device (see Supplementary Fig S2). In each measurement, the CMOS image sensor was configured to take 60 frames of time-lapse holographic images. Fifty different measurements were conducted on each type of pollen within a period of a month, without any need to replace the virtual impactor. This capability to conduct long-time experiments is a unique feature enabled by the virtual impactor, which does not immobilize the particles or create contamination on the imaging window.

These time-lapse holographic images monitor the particles passing through the imaging window above the image sensor. A flying particulate matter can be easily identified from the time-lapse holograms captured by the portable device (Fig. 3); without loss of generality, the triplicate holographic patterns of a flowing particle of interest appear in the second frame of any consecutive three-frame holograms: $H(t_0)$, $H(t_0 + \Delta t)$ and $H(t_0 + 2\Delta t)$. A differential hologram $H_d$ can be calculated from these three successive CMOS frames (Fig. 3), i.e.,

$$H_d = 2H(t_0 + \Delta t) - \big(H(t_0 + 2\Delta t) + H(t_0)\big). \tag{1}$$

Note that $\Delta t$=285.7 ms and should not be confused with the interval between two successive pulses ($\delta t = 699 \mu s$); the latter creates multiple holograms of each flowing particle on a single lensfree image frame, whereas $\Delta t$ is determined by the frame rate of the CMOS imager (3.5 fps). After this differential calculation, the flying particles of interest that only appear in the frame $H(t_0 + \Delta t)$ present a lower intensity level compared to the background. A threshold was applied to $H_d$ for localizing each flowing particle, and the resulting image patches with the detected particles were cropped, each with a size of 256×256 pixels.

The microscopic image of the flowing pollen in each region of interest (ROI) was reconstructed by digitally propagating the raw hologram using the angular spectrum method to its focal plane, estimated by an autofocusing algorithm[49]. For all the six pollen species used in this work, the raw triplicate holograms of some representative particles and their back-propagated images are illustrated in Fig. 4. The back-propagated holograms of each pollen, without any motion blur or rolling shutter artifact, showed a good reconstruction fidelity. Importantly, the three holographic replicas of the same particle were not identical, which indicates that the particles have free rotation while flowing within the air stream. These images of each pollen from different perspectives provided richer information about the pollen, which proved important for more accurate classification of their type, as will be reported next.

An image dataset was established from the experimental data captured using our device, containing ~6000 ROIs for training and validation, and ~900 ROIs for blind testing. A deep neural network based on DenseNet201[50] was trained to classify the type of pollen particles. During the training, the network treated each ROI as an *independent* particle. In other words, the relation between the holographic replicas of each flowing particle was intentionally ignored to increase the robustness of our classification system. To utilize the volumetric information provided by lensfree holographic imaging, the network was trained with both the real and imaginary channels of each ROI propagated to five axial locations: its focus plane and 50μm and 100μm above and below the focus plane. After the training phase, the deep neural network achieved a blind testing accuracy of 90.48% in classifying all the individual ROIs containing 6 different types of pollen. The corresponding confusion matrix is displayed in Fig. 5(a). Using the additional information (with different perspectives of the pollen) available in triplicate holographic images of each flowing particle helps improve the final classification accuracy. The ROIs belonging to the same particle were first located and grouped together. To utilize this additional source of information, we devised a majority voting scheme applied to the labels of the three successive ROIs (corresponding to three holographic replicas of the same particle), which were independently classified using the same trained network. The voting winner (with 2 or more votes) was selected as the final predicted label (pollen type) for all three replicas of each flowing pollen particle, which increased the blind testing accuracy to 92.91% (see the confusion matrix in Fig. 5(b)).

## Discussion

The presented particle classification scheme using triplicate holographic images has a unique advantage since it permits the visualization of the same particle from different perspectives during its flow within the virtual impactor. The impact of this capability can be better seen in the classification of pine pollens. In general, pine pollens have a unique feature with wing-like bladders (see the fourth column in Fig. 4 as an example). During its flow inside the virtual impactor channel, there is a chance that the captured hologram of a pine pollen only reflects its main body, with the bladders hidden behind it, which makes it resemble a wheat pollen of similar size (see Supplementary Fig. S3 as an example). In fact, because of this, the trained neural network misclassified 22.97% of pine pollens as wheat using a single holographic image. The classification accuracy of pine pollens was improved using the triplicate holographic images through a majority voting process, reducing the error rate to 18.57%. Furthermore, if we relax our voting rule for pine pollens such that all the particles in triplicate holographic images will be labeled as pine pollens if at least one of those holographic images was classified as pine, the error rate can further drop to 17.14%.

The designed virtual impactor device with lens-free holographic imaging also presents a unique feature of volumetric sensing of flowing particulate matter. It allows us to image flowing particles distributed inside a large volume (25 mm3) and records the 3D information about the particles through holography. Harvesting this 3D information, in this work, we demonstrated a deep neural network that utilizes the complex-valued images of each flowing particle at five different axial locations, each of which has different phase and amplitude profiles, reflecting the unique 3D refractive index information of the particle. To shed more light on the classification advantages brought by this volumetric sensing approach, we further trained another deep neural network taking only the auto-focused images, i.e., from a single axial location, while keeping the architecture and training parameters the same as before. This network that only used the complex-valued object field from a single axial plane achieved a worse classification accuracy of 85.23%, which indicates the advantages of using multiple complex fields at different axial planes for each flowing particle.

Generally, the quality of an inline holographic image after a simple auto-focusing step suffers from the twin-image artifact unless phase recovery is applied to it. In this study, however, since the distance between the samples and the CMOS image sensor is relatively large (on average ~ 2 mm), we did not use phase retrieval and the twin image artifact did not constitute an obstacle to the accurate classification of pollen particles. In case a more clear microscopic image of each flowing particle is desired, iterative or neural network-based phase recovery algorithms[51–54] can be used to remove the twin-image artifacts, revealing increased contrast and SNR for each particle. In addition, deep neural networks can also be used for auto-focusing and phase recovery at the same time[53,55,56]. The inclusion of these additional processing steps might further improve the performance of our virtual impactor-based label-free bio-aerosol detection and classification device.

## Conclusion

In conclusion, we presented a novel device for label-free bio-aerosol detection using a virtual impactor and computational imaging. Pollen particles in the air were slowed down and concentrated in the middle channel of a virtual impactor. Pulsed illumination was used to form triplicate holographic patterns of the same particle on a single frame. The volumetric sensing provided by holographic multi-shot imaging of the same flowing particle brought unique features to this device, using which a deep neural network achieved 92.91% accuracy in classifying different types of pollens. The mobile device prototype costs ~$920 and weighs ~700g, which can be further reduced in mass production, providing a cost-effective and portable solution to long-term unattended personalized bio-aerosol monitoring.

# Methods

## Virtual impactor design and fabrication

The cut-off diameter $d_{50}$ of a virtual impactor can be estimated using the flow velocity and the virtual impactor geometry[26]:

$$d_{50} = \sqrt{\frac{9\eta W^2 L (Stk_{50})}{\rho_p Q}} - 0.0078 \times 10^{-6} \qquad (2)$$

where $\eta$ is the air viscosity. $W$ and $L$ denote the width and length of the impaction nozzle (the junction of the middle and by-pass channels). $Stk_{50}$ is the Stokes number of the particles with 50% collection efficiency, $\rho_p$ is the density of the particle, and $Q$ is the flow rate. In designing the virtual impactor, the flow rate Q was first empirically chosen, and the dimension of the virtual impactor nozzle was optimized using the FEM solver in COMSOL (see Fig. 2(a)). The virtual impactor channel was fabricated using a 3D printer (Objet30 Pro, Stratasys Inc.) with a light-blocking material. The region designed for the imaging window was fabricated as open holes during the 3D printing. Coverslip glasses were used to seal them and form an air-tight channel.

## Camera exposure settings and illumination pulse synchronization

The CMOS sensor (IDS-UI-3592LE-C-VU, 4912 x 3684 pixels, pixel pitch 1.25μm) used in this work operates based on a rolling shutter with a global release feature. The camera sequentially turns on all 4912 rows of pixels to start collecting photons, from top to bottom, and sequentially turns them off. The time between the bottom row of the pixels to start detecting photons and the top row to stop detecting photons was set to be 2600 μs. In other words, all pixels on the CMOS sensor collect photons during this 2600 μs period. A high voltage level is provided when all pixels are turned on. This signal triggers a single pulse with a duration of 1600 μs from a re-triggerable monostable multivibrator (74LS123, Texas Instrument Inc.). This single pulse is further coupled with a pulse train generated by a 555 timer (LMC555CN, Texas Instrument Inc.) with a pulse width of 6.9 μs and a period of 707 μs. Three pulses (each having 6.9 μs) are generated and sent to an LED controller (TLC5917, Texas Instrument Inc.), which injects 120 mA current into the laser diode (PLT5 510, OSRAM Opto Semiconductors GmbH).

**Neural network training**

The network used in this work was adapted from DenseNet201[50], with the channel number of the first convolutional layer tuned to match the input image channels. In the network training, each ROI was first randomly cropped to have a size of 224×224 pixels. Data augmentation, including random flipping and rotation, was consequently applied to the images. Finally, before being fed into the networks, each input image was processed with a Gaussian blur to remove salt and pepper noise caused by the short exposure under each pulsed illumination (6.9 μs). A softmax cross entropy loss was calculated using the network predicted class scores and the ground truth pollen species as:

$$\mathcal{L}_I = -\sum_{c=1}^{6} g_c \cdot \log\left(\frac{\exp(s_c)}{\sum_{c\prime=1}^{6} \exp(s_{c\prime})}\right) \quad (3),$$

where $s_c$ denotes the predicted class score for the $c^{\text{th}}$ class, and $g_c$ denotes the $c^{\text{th}}$ entry of the ground truth label vector. The network parameters were optimized using an Adam optimizer, with the learning rate set to be $1\times10^{-4}$ at the beginning and tuned using a cosine annealing schedule. The network was trained using a desktop computer with a Ryzen 9 3950X central processing unit (AMD Inc.) and an RTX 2080Ti graphic processing unit (GPU, NVidia Inc.) with 64 GB of memory, running on Windows 10 (Microsoft Inc.). The typical training time with 200 epochs is ~5 hours.

# Figures

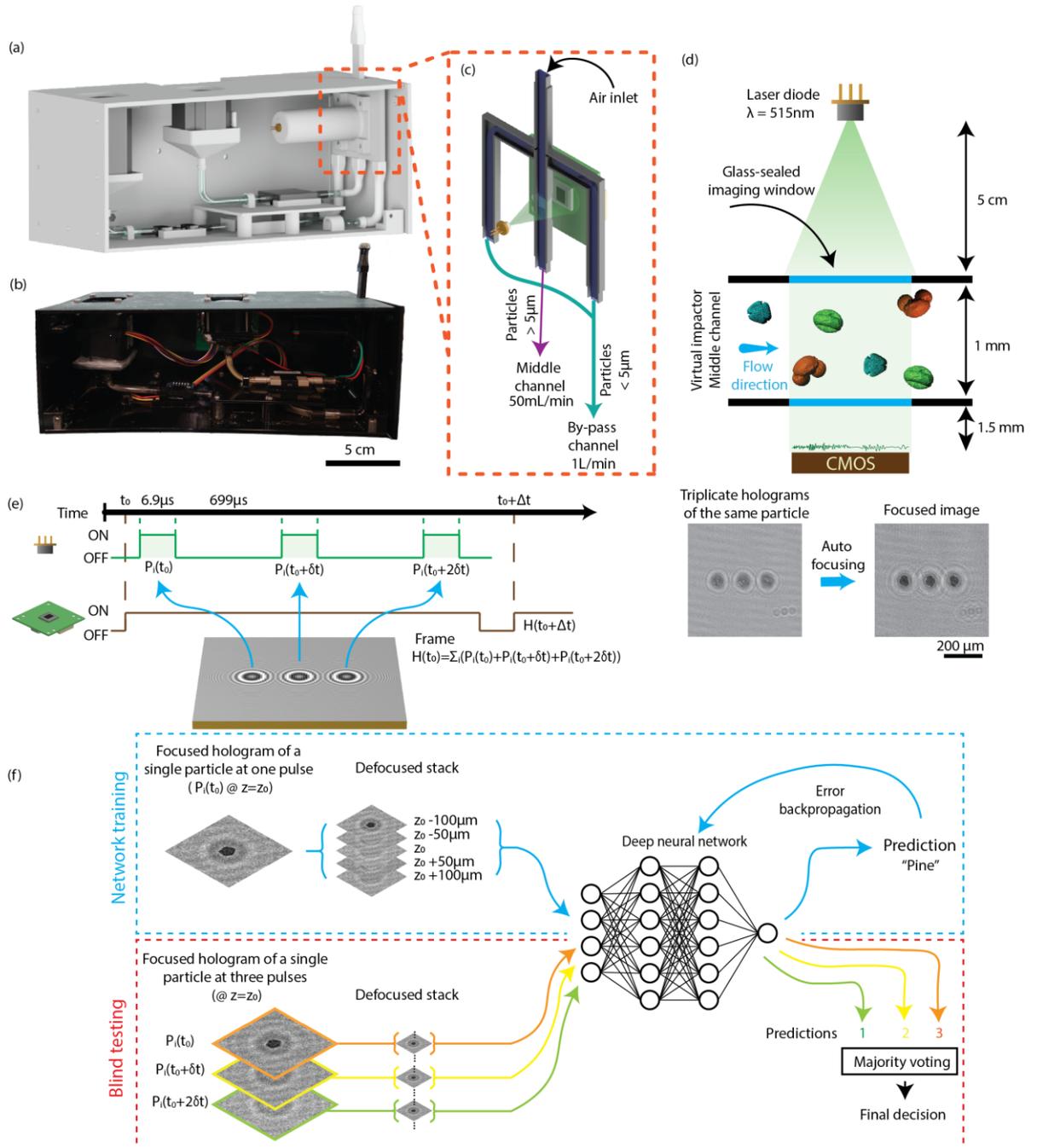

**Figure 1.** The virtual impactor-based label-free bio-aerosol detection device. (a) 3D computer-aided-design (CAD) drawing of the device. (b) Photograph of the prototype. Major parts of the device were 3D printed. (c) Schematic drawing of the virtual impactor channel used to concentrate the flowing aerosols and the lens-free imaging set-up. (d) Schematic drawing of the lens-free imaging system. A laser diode illuminated the particles flying through the imaging window with pulsed illumination. (e) Three pulses formed a pulse train. Particles cast lens-free inline holograms on a CMOS image sensor with each pulse. Triplicate holographic patterns are digitally integrated by the CMOS sensor on each frame. (f) A deep neural network was trained to classify pollen species. FOVs containing individual pollen particles were first auto-focused. The focused hologram was intentionally defocused for $\pm 50$ and $\pm 100 \mu m$ in the axial direction. Both the real and imaginary channels of the focused and defocused holograms formed an image stack to train the network. In the blind testing stage, a majority voting was applied to the labels inferred using the triplicate holographic patterns of each flowing particle.

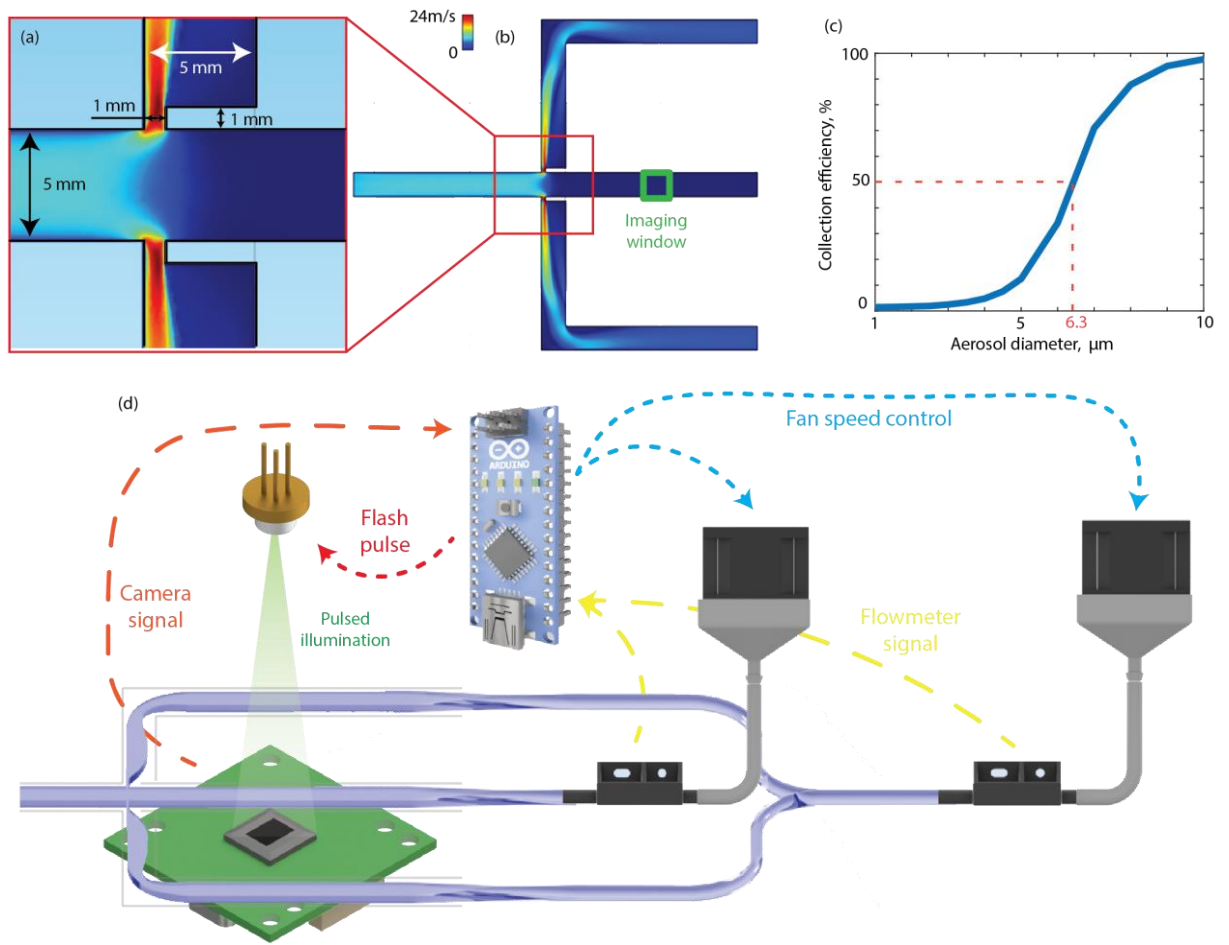

**Figure 2.** Design of the virtual impactor-based system. The geometrical design of the virtual impactor used to concentrate bio-aerosols was optimized using FEM simulations. (a-b) The flow field inside the channel. (c) Simulated particle collection efficiency in the middle channel of the virtual impactor. (d) Schematic drawing of the controlling circuit for the portable device. On the left, the circuit receives the signal from the CMOS image sensor and fires a pulse train. On the right, the circuit measures the flow rate inside the virtual impactor channel and uses it to adjust the fan power using a PID controller.

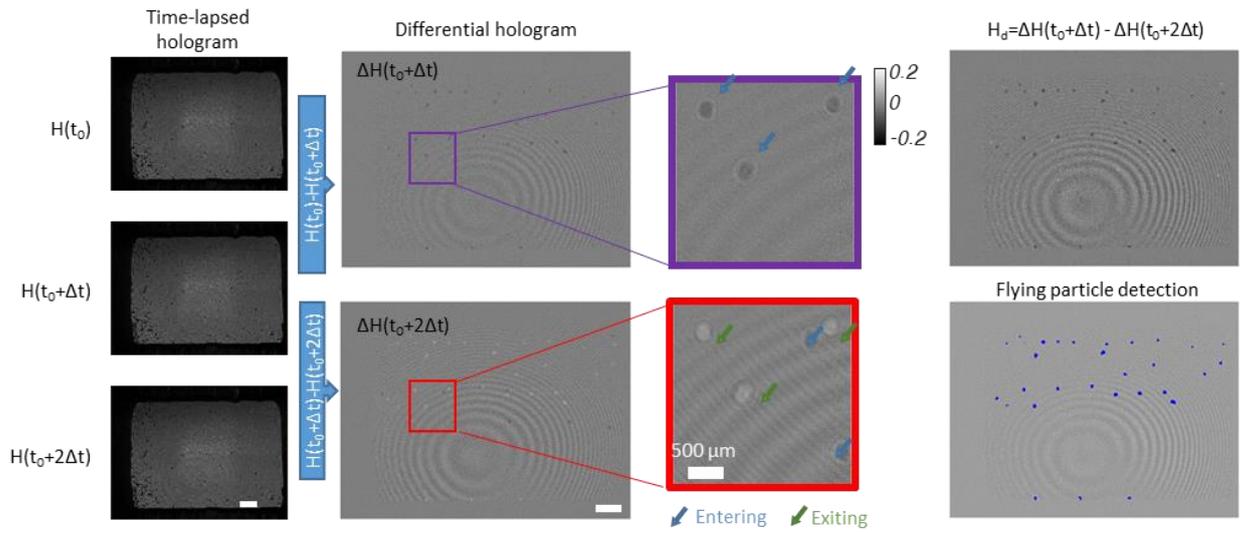

**Figure 3.** Flowing particle detection and localization using differential holograms. Differential holograms (see Eq. 1) were calculated using three successive holograms $H(t_0)$, $H(t_0 + \Delta t)$ and $H(t_0 + 2\Delta t)$. Note that $\Delta t$=285.7 ms and should not be confused with $\delta t = 699 \mu s$ which is the interval between two successive pulses. One of these holographic images contains the entire triplicate hologram set of a flowing particle within the imaging window. Flying particles present lower intensity levels in the differential hologram $H_d$ and a threshold was used to localize the flowing particles.

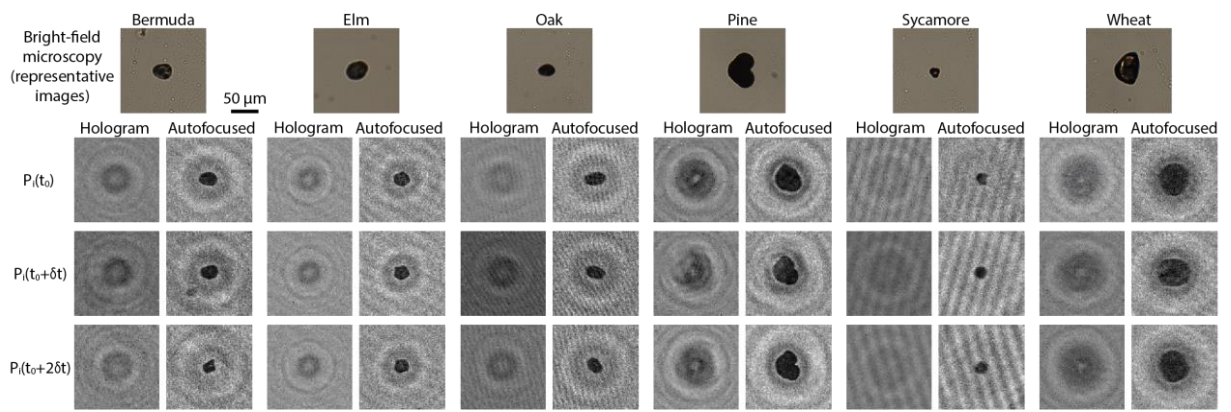

**Figure 4.** Triplicate holographic patterns of the same flowing particles under pulsed illumination; each pulse duration is 6.9 $\mu s$ and $\delta t = 699 \mu s$. Top row: a representative pollen particle of each species imaged under a bright-field microscope. Bottom three rows: triplicate holographic patterns of a different pollen particle of each species captured during our experiments; their autofocused images are also shown to the right of the corresponding lensfree hologram.

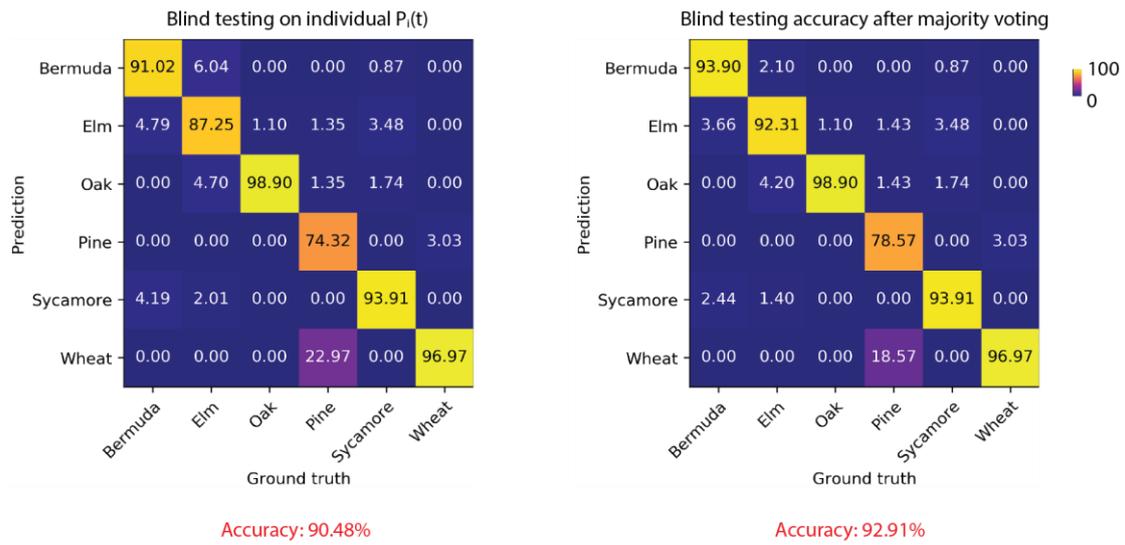

**Figure 5.** Confusion matrices of the classification deep neural network. (Left) In the blind testing stage, each image FOV is separately/individually classified, ignoring triplicate holograms of each flowing particle. (Right) A majority voting was applied to the three labels predicted by using the triplicate holographic patterns of each flowing particle. The final classification accuracy with majority voting improved to 92.91%.